\documentclass[conference]{IEEEtran}
\IEEEoverridecommandlockouts
\usepackage{cite}
\usepackage{amsmath,amssymb,amsfonts}
\usepackage{algorithmic}
\usepackage{graphicx}
\usepackage{textcomp}
\usepackage{xcolor}
\usepackage{url}
\def\BibTeX{{\rm B\kern-.05em{\sc i\kern-.025em b}\kern-.08em
    T\kern-.1667em\lower.7ex\hbox{E}\kern-.125emX}}
\begin{document}

\title{Stock Market Directional Bias Prediction Using ML Algorithms
}

\author{\IEEEauthorblockN{Ryan Chipwanya}
\IEEEauthorblockA{\textit{Academy of Computer Science and Software Engineering} \\
\textit{University of Johannesburg}\\
Johannesburg, South Africa \\
219027938@student.uj.ac.za}
}

\maketitle

\begin{abstract}
The stock market has been established since the 13th century, but in the current epoch of time, it is substantially more practicable to anticipate the stock market than it was at any other point in time due to the tools and data that are available for both traditional and algorithmic trading. There are many different machine learning models that can do time-series forecasting in the context of machine learning. These models can be used to anticipate the future prices of assets and/or the directional bias of assets. In this study, we examine and contrast the effectiveness of three different machine learning algorithms—namely, logistic regression, decision tree, and random forest—to forecast the movement of the assets traded on the Japanese stock market. In addition, the models are compared to a feed forward deep neural network, and it is found that all of the models consistently reach above 50\% in directional bias forecasting for the stock market. The results of our study contribute to a better understanding of the complexity involved in stock market forecasting and give insight on the possible role that machine learning could play in this context. 

\end{abstract}

\begin{IEEEkeywords}
Machine Learning,  Stock Market, Prediction, Classification
\end{IEEEkeywords}

\section{Introduction}
Forecasting the stock market and carrying out algorithmic trading both benefit significantly from the application of machine learning (ML). A skill that can be acquired, predicting the stock market entails learning and utilising information and resources on both fundamental and technical analytical techniques in order to estimate the future price of an asset [1]. This talent can be acquired via practise.

Traditional trading methodologies, on the other hand, introduce an increased probability of errors in accurate predictions. These errors can be caused by human emotions (such as fear and greed) that drive impulsive trading behaviour in high-volume market conditions, unprovoked fundamental news events, and a lack of necessary skills to adequately forecast assets [4].

The availability of essential data, including as news, prices, and indicators for critical analysis and forecasting [5,6], has made it significantly simpler throughout the course of human history to foresee the behaviour of various markets, and this trend continues into the present day.

Studies have continued to address the problems in exactly predicting future stock prices; yet, they have also proceeded to reach promising findings in asset forecasting, evaluated using ML and Deep Learning (DL) algorithms, proving to attain above 80 percent accuracy in forecasting stocks [1,2]. Despite the fact that studies have continued to discuss the difficulties in precisely predicting future stock prices. ML Algorithms have made it possible to estimate future prices or directional biases for stocks and other asset classes with significant outcomes [1,2].

The results of this study provide valuable insights into the investigation of a variety of ML and DL approaches to stock market forecasting. In this work, we investigate how well ML Algorithms like Logistic Regression, Decision Trees, and Random Forest perform when it comes to forecasting the directional bias in assets that are listed on the Japanese Stock Exchange, and we compare their results to those of the most advanced DL models available. It is believed that this research will help to a greater understanding of the significance of ML and DL algorithms in the process of predicting stock market prices.

\section{Method}
\subsection{ML Models}
In order to evaluate the efficacy of stock market forecasting, we will set up the task as a binary classification one, in which we will predict whether the movement of the stock for the day will be "up" = 1 or "down" = 0. Thawornwong and Enke's research demonstrated that directional predictions for stocks perform better than exact numerical predictions [7]. In addition, we create three machine learning models, which are referred to as a Logistic Regressor (LR), a Decision Tree (DT), and a Random Forest (RF). We will utilise binary entropy and entropy for the Logistic Regressor as well as accuracy as a metric to evaluate the performance of the LR and DT + RF models, respectively, in order to calculate the impurity.

\subsection{The Data set}
The Kaggle JPX Tokyo Stock Exchange Prediction Competition, which was organised by the Japanese Exchange Group [8] will serve as the data set that we will be utilising in this study. In addition, for the purpose of assessment, we will apply a filter to the data set, extract the price data for Sony, and then remove the columns labelled "Date," "Open," "High," "Low," and "Close," as well as the "Volume" column. In addition to this, we are going to make two more columns and label them "Next" and "Target." The "Next" column displays the "Close" price for the following trading day, while the "Target" column is used to classify whether the movement will be an increase or a decrease, shown by the numbers 1 and 0 accordingly. The "Target" column is the result of applying a comparison operator to the daily data, which compares the prices at the end of the current day to those of the following day. "Date", "Open", "High", "Low", "Close", and "Volume" will be the features that are retrieved and used for training in order to properly anticipate this time-series forecasting domain.

\begin{figure}[htbp]
\centerline{\includegraphics[width=1.2\columnwidth]{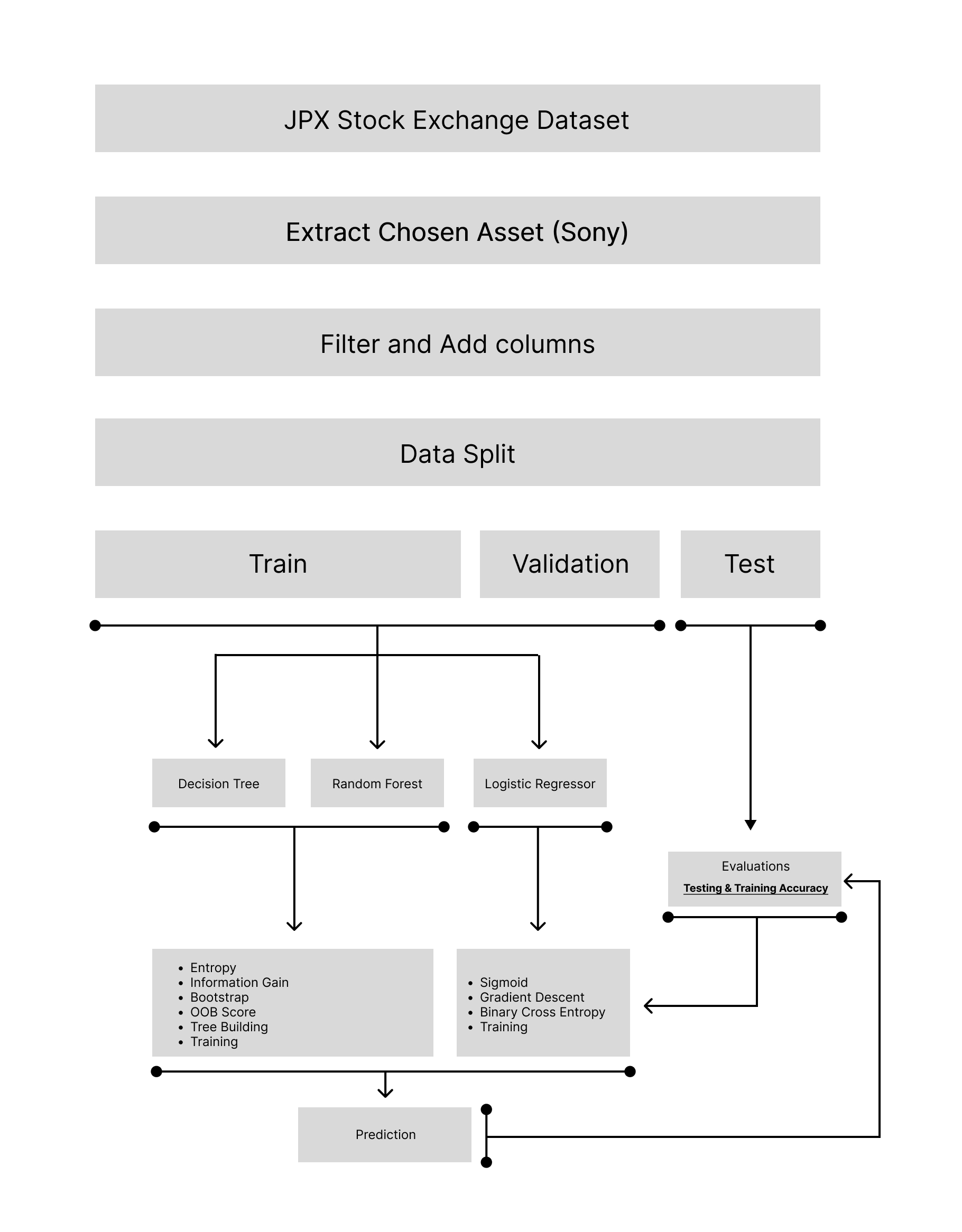}}
\caption{The proposed framework}
\label{fig}
\end{figure}

\section{Results}
The outcomes of each machine learning model are summarised in the tables that follow. The model setup for the logistic regression model can be seen in Table 1. This configuration exhibits the implemented functions, the training information, and the model assessment metrics. Table 2 provides an overview of the DT and RF classifiers, including the parameters, types of sampling, and model evaluation metrics that are specific to each classifier. Table 3 provides a comparison of the various models' performance as well as their overall cost.

In addition, in order to evaluate how well our ML models perform in comparison to those of our DL model, a Feed forward Neural Network (FNN) was developed throughout the course of the research using the TensorFlow library and given the same task as the ML models. In order to fine-tune the FNN architecture's predictive capabilities, some hyper parameters were adjusted throughout the configuration process. An input layer that consisted of five neurons was included in the model. These neurons corresponded to the features that were selected for the task. These features included 'Close,' 'Volume,' 'Open,' 'High,' and 'Low.' After that, two hidden layers with a total of 128 and 64 neurons were added to the model in order to give it the ability to recognise detailed patterns in the financial data. In the output layer, which was designed specifically for binary classification, a single neuron equipped with the sigmoid activation function was used to determine whether or not the stock price will go up. Model optimisation was accomplished through the use of the Adam optimizer, and the binary cross-entropy loss function was selected as the optimal option given the characteristics of the binary classification challenge. Table 4 displays the architecture of the FNN. The FNN was put through a strenuous training routine that lasted for ten epochs and included a batch size of 32 in addition to validation monitoring. The performance of the model was evaluated using a specific test data set, and the results showed that the model had a test loss of roughly 0.68 and a test accuracy of 59\%. The F1 score, which exhibited a balanced performance with a score of 0.74, demonstrated that the model was proficient in both precision and recall. However, it is important to note that the confusion matrix highlighted a substantial class imbalance. This indicated that the model consistently identified all occurrences as the positive class, which highlights the necessity for additional analysis and future model changes. In contrast, the accuracy and F1 score have similar performance to our baseline and state-of-the-art models, which have proven that adopting a DL or ML model does not make a substantial difference in the results.

\subsection{Evaluation}

In the course of our empirical research, we predicted the direction of the bias by applying three different machine learning models: logistic regression (LR), decision tree (DT), and random forest (RF). Our LR model scored a remarkable accuracy of 55\%, which indicates that it is able to accurately estimate directional trends in slightly more than half of the situations. The LR model was surpassed by the baseline DT model, which had an accuracy rate of 59\%. This finding demonstrates that it has more advanced skills for recognising data patterns than LR does. In conclusion, the high-performing RF classifier had an accuracy of 63\%, demonstrating the benefits of using ensemble methods to improve prediction accuracy.

\begin{table}[!t]
\renewcommand{\arraystretch}{1.3}
\caption{Experimental Setup for Logistic Regression Model}
\label{tab:logistic_regression_setup}
\centering
\begin{tabular}{p{0.2\linewidth}p{0.6\linewidth}}
\hline
\textbf{LR Model} & \textbf{Description}\\
\hline
Functions & \begin{itemize}
  \item \texttt{\raggedright  sigmoid(X, theta)}: Computes the sigmoid function for logistic regression.
  \item \texttt{\raggedright  entropy(X, y, theta)}
\raggedright   \item \texttt{\raggedright  logistic\_regression(X, y, alpha, epochs)}: Trains the logistic regression model using gradient descent. It takes learning rate ($\alpha$) and the number of training epochs (epochs) as hyper parameters.
\end{itemize} \\
\hline
\raggedright  Training Details & \begin{itemize}
  \item Trained the logistic regression model using the training data.
  \item Learning Rate ($\alpha$): 0.01
  \item Number of Training Epochs (epochs): 1000
  \item Model's parameters ($\theta$) and cost history are saved.
\end{itemize} \\
\hline
\textbf{Model Evaluation} & \\
\hline
Model Testing & \begin{itemize}
\raggedright   \item Made predictions on the validation set using the trained logistic regression model.
\end{itemize} \\
\hline
\raggedright Performance Metric & \begin{itemize}
  \item Evaluated the model's performance using accuracy as the evaluation metric.
  \item The accuracy score was calculated by comparing the model's predictions to the true target values.
\end{itemize} \\
\hline
\raggedright \textbf{Experimental Configuration} & \\
\hline
\raggedright Learning Rate ($\alpha$) & 0.01 \\
\hline
\raggedright Number of Training Epochs (epochs) & 1000 \\
\hline
\end{tabular}
\end{table}

\begin{table}[!t]
\renewcommand{\arraystretch}{1.3}
\caption{Experimental Setup Details for DT \& RF}
\label{tab:experimental-setup-dt-rf}
\centering
\begin{tabular}{p{0.2\linewidth}p{0.7\linewidth}}
\hline
\textbf{Component} & \textbf{Description} \\
\hline
\textbf{DT \& RF Parameters} & 
\begin{itemize}
  \item Number of Trees (Estimators): 250
  \item Maximum Features: 5
  \item Maximum Depth: 100
  \item Minimum Samples Split: 100
\end{itemize} \\
\hline
\textbf{Sampling} & Bootstrap sampling to create multiple subsets of the training data, used to train individual decision trees. \\
\hline
\textbf{Error Estimation} & 
The Out-of-Bag (OOB) error is calculated by measuring mis-classifications on data points not included in the bootstrap sample. \\
\hline
\textbf{Model Evaluation} & The validation accuracy of the model by comparing its predictions to the true target values. \\
\hline
\end{tabular}
\end{table}

\begin{table}[!t]
\renewcommand{\arraystretch}{1.3}
\caption{Experimental Results Validation Accuracy \& F1 Score}
\label{tab:experimental-results}
\centering
\small 
\resizebox{\columnwidth}{!}{ 
\begin{tabular}{|p{2.0cm}|p{3.0cm}|p{3.0cm}|p{3.0cm}|}
\hline
\textbf{Model} & \textbf{Accuracy} & \textbf{F1 Score} \\ \hline
LR  & 0.55 & 0.71 \\ \hline
DT  & 0.59 & 0.74 \\ \hline
RF  & 0.63 & 0.74 \\ \hline
\end{tabular}
}
\end{table}

\begin{table}[!t]
\renewcommand{\arraystretch}{1.3}
\caption{Experimental Setup for Feed forward Neural Network Model}
\label{tab:feedforward_nn_setup}
\centering
\begin{tabular}{p{0.20\linewidth}p{0.7\linewidth}}
\hline
\raggedright \textbf{Feed forward Neural Network Model} & \textbf{Description}\\
\hline
Architecture & \begin{itemize}
  \item Input Layer: $5$ input features (Close, Volume, Open, High, Low).
  \item Hidden Layers: Two hidden layers with $128$ and $64$ neurons, respectively.
  \item Output Layer: $1$ neuron with a sigmoid activation function.
\end{itemize} \\
\hline
\raggedright Training Details & \begin{itemize}
  \item Trained the feed forward neural network model using the training data.
  \item Optimizer: Adam optimizer.
  \item Loss Function: Binary cross-entropy loss.
  \item Number of Training Epochs: $10$
  \item Batch Size: $32$
  \item Validation Split: $20\%$ of the training data.
\end{itemize} \\
\hline
Model Evaluation & \\
\hline
Model Testing & \begin{itemize}
  \item Made predictions on the test set using the trained feedforward neural network model.
\end{itemize} \\
\hline
\raggedright Performance Metrics & \begin{itemize}
  \item Evaluated the model's performance using the following metrics:
    \begin{itemize}
      \item Test Loss: Binary cross-entropy loss on the test set.
      \item Test Accuracy: Accuracy of the model on the test set.
      \item F1 Score: F1 score for binary classification on the test set.
      \item Confusion Matrix: Confusion matrix on the test set.
    \end{itemize}
\end{itemize} \\
\hline
\raggedright Experimental Configuration & \\
\hline
\raggedright \raggedright Learning Rate \& Adam optimizer default (adaptive learning rate) & \\
\hline
\raggedright \raggedright \raggedright Number of Training Epochs & 10 \\
\hline
\end{tabular}
\end{table}

\section{Discussion}
\subsection{Summary of key findings}

In this study we applied machine learning methods LR,DT and RF to predict the directional bias in assets listed on the Japanese Stock Exchange [8]. We discussed our primary objective which was to comparatively evaluate the performance of the models in the context of stock market predictions and achieved plausible results. 

Our LR model being labelled as our Naive model, managed to achieve an accuracy of 55\% which indicates an impressive naive performance which paves the way to compare to more technically complex ML algorithms such as our baseline DT and RF models. The DT classifier, labelled as our baseline model achieved a validation accuracy of 59\% which showed a mammoth increase in validation accuracy compared to the LR model and lastly our state-of-the-art (SOTA) RF model achieved performance of 63\% displaying the added advantage of the ensemble of DT's. 

In the context of existing literature, our models prove to align with existing research in ML algorithms for stock market prediction in the discussion of accuracy, in particular Zhong and Enke's study suggested that for stock market binary classification , ML models may generally achieve around a 60\% accuracy which is in margin of our baseline and SOTA models [9]. Furthermore, DL models such as Artificial Neural Network and Deep Neural Network  experimented in the same study by Zhong and Enke , achieved results of 58.6\% and 59.9\% , in-line with out DT and RF classifiers. 

\section{Conclusion}

In conclusion, the findings of our research show that machine learning algorithms are capable of accurately predicting the direction in which assets traded on the Japanese stock market will tend to move. According to the findings of our investigation, a number of models, such as the Logistic Regressor, Decision Tree, Random Forest Classifier, and the Deep Learning Feed Forward Neural Network (FFNN) model, routinely attain accuracy rates that are higher than 50\%. These findings have important repercussions for the study of financial data and the development of investment strategies.

However, it is essential to recognise the limitations of our study, specifically the lack of advanced metrics and trading performance data. This absence raises doubts regarding the profitability and return on investment (ROI) of the models in question. Due to the unpredictability of the market, the financial risks connected with adopting predictions generated by machine learning for trading portfolios remain substantial.

In order to improve on these findings, future research should investigate the possibility of including technical analysis indicators and use advanced deep learning models suitable for time-series forecasting, such as Long Short-Term Memory (LSTM) and Recurrent Neural Networks (RNN). For instance, Zaini et als paper found that using technical indicators with a LR model achieve a classification accuracy of 86\% [10]. From this, it is possible that this will result in an improvement in the accuracy of directional bias predictions in stock markets.

Our research contributes to the understanding of the complexity of accurately predicting the behaviour of the stock market and paves the way for additional investigation of methods, tools, and indicators that have the potential to enhance the effectiveness of machine learning models. The application of machine learning to the task of stock market forecasting has the potential to have far-reaching repercussions not only for investors but also for the financial industry as a whole as the field of financial technology continues to advance. This research provides a basis for future studies that aim to solve the obstacles and uncertainties involved with financial prediction using machine learning models. These investigations will be conducted in the future.


\begin{thebibliography}{00}
\bibitem{b1}
I. Parmar et al.,
\emph{Stock Market Prediction Using Machine Learning},
2018 First International Conference on Secure Cyber Computing and Communication (ICSCCC), Jalandhar, India, 2018,
pp. 574-576,
doi: 10.1109/ICSCCC.2018.8703332.

\bibitem{b2}
P. S and V. P. R,
\emph{Stock Price Prediction using Machine Learning and Deep Learning},
2021 IEEE Mysore Sub Section International Conference (MysuruCon), Hassan, India, 2021,
pp. 660-664,
doi: 10.1109/MysuruCon52639.2021.9641664.

\bibitem{b3}
L. Mathanprasad and M. Gunasekaran,
\emph{Analysing the Trend of Stock Market and Evaluate the performance of Market Prediction using Machine Learning Approach},
2022 International Conference on Advances in Computing, Communication and Applied Informatics (ACCAI), Chennai, India, 2022,
pp. 1-9,
doi: 10.1109/ACCAI53970.2022.9752616.

\bibitem{b4}
A. W. Lo, D. V. Repin, and B. N. Steenbarger,
\emph{Fear and Greed in Financial Markets: A Clinical Study of Day-Traders},
\emph{The American Economic Review},
vol. 95, no. 2, 2005,
pp. 352--359,
\url{http://www.jstor.org/stable/4132846}.

\bibitem{b5}
\raggedright I. Hwang, A Brief History of the Stock Market, June 15, 2023, Accessed September 28, 2023,
\url{https://www.sofi.com/learn/content/history-of-the-stock-market/}.

\bibitem{b6}
Y. Han, Y. Liu, G. Zhou, and Y. Zhu,
\emph{Technical Analysis in the Stock Market: A Review},
SSRN,
May 21, 2021,
Available at SSRN: \url{https://ssrn.com/abstract=3850494}
or \url{http://dx.doi.org/10.2139/ssrn.3850494}.

\bibitem{b7}
S. Thawornwong and D. Enke,
\emph{The adaptive selection of financial and economic variables for use with artificial neural networks},
\emph{Neurocomputing},
vol. 56, 2004,
pp. 205--232.

\bibitem{b8}
\raggedright \raggedright A. Sugiyama, C. Hio(Alpaca), E. Kaji, n-onishi, s-meitoma - JPX, S. Takato, and T. Kitayama(Alpaca),
\emph{JPX Tokyo Stock Exchange Prediction},
Kaggle, 2022,
\url{https://kaggle.com/competitions/jpx-tokyo-stock-exchange-prediction}.

\bibitem{b9}
X. Zhong and D. Enke,
\emph{Predicting the daily return direction of the stock market using hybrid machine learning algorithms},
\emph{Financial Innovation},
vol. 5, 2019,
p. 24,
doi: 10.1186/s40854-019-0138-0.

\bibitem{b10}
\raggedright Jamili Zaini, Bahtiar, Rosnalini Mansor, Norhayati Yusof, and Beh Hui Sang. 2020.“Classify Stock Market Movement Based on Technical Analysis Indicators Using Logistic Regression”. Journal of Advanced Research in Business and Management Studies 14 (1):35-41. 
\bibitem{b11}
Deep learning for financial time series
forecast fusion and optimal portfolio
rebalancing, S Laher, A Paskaramoorthy, TL Van Zyl, 2021 IEEE 24th International Conference on Information Fusion (FUSION), 1-8
\bibitem{b12}
Parden: Surrogate assisted hyper-
parameter optimisation for portfolio selection, TL van Zyl, M Woolway, A Paskaramoorthy, 2021, 8th international conference on soft
computing \& machine intelligence

\end{thebibliography}
\end{document}